\documentstyle[12pt]{article}
\begin{document}
\begin{center}
{\Large \bf{A constant equation of state for quintessence~?}}
\vspace*{7mm} \\
Elisa Di Pietro and Jacques Demaret
\vspace*{10mm} \\
Institut d'Astrophysique et de G\'eophysique, \\ \noindent
Universit\'e de Li\`ege, \\ \noindent
Avenue de Cointe, 5 , B-4000 
Li\`ege, Belgium \vspace*{2mm} \\ \noindent
dipietro@astro.ulg.ac.be
\vspace*{15mm} \\
\end{center}
\begin{abstract}
\noindent
Quintessence is often invoked to explain the universe's acceleration suggested
by type Ia supernovae observations. The aim of this letter is to study the 
validity of using a constant equation of state for quintessence models. We shall 
show that this hypothesis strongly constraint the form of the scalar 
potential.
\end{abstract}
.\vspace*{15mm} 
\newpage 
\section{Introduction} 
The Standard Cosmological Model (SCM) can only describe decelerated universe
models and thus cannot reproduce the observations coming from the recent type 
Ia supernovae, CMB anisotropies, ... in favour of a presently accelerated 
universe (see e.g. \cite{sneia}). But, as the SCM can give a satisfactory 
explanation to other observational properties of the present Universe (e.g. 
primordial nucleosynthesis, extragalactic sources'redshifts, cosmic microwave 
radiation), the tendency is to consider the SCM as incomplete rather than 
incorrect.

The SCM can be transformed in an accelerated model by adding a new ingredient
which behaves as a fluid with a negative pressure. The oldest and most 
studied candidate for this missing component is the cosmological constant 
$\Lambda$ which is equivalent to a perfect fluid with constant density and
pressure related by the equation of state $p = - \rho$ \cite{constant}.
However, this does not constitute the only possibility: among all the other
candidates, this missing energy can be associated with a dynamical
time-dependent and spatially (in)homogeneous scalar field $\phi$ evolving
slowly down its potential $V(\phi)$. The resulting cosmological models are
known as {\em quintessence models} \cite{quintessence}. In these models, the
scalar field can be interpreted as a fluid with a negative pressure given
by $p=w\,\rho$ ($-1\leq w<0$). Quintessence allows a wide range of models
including a constant or a time-varying {\em w}. Nevertheless some of these 
models admit an attractive property, called {\em tracker solution}
\cite{tracker}, that permits to solve the cosmic coincidence
problem\footnote{``Cosmic coincidence problem'' refers to an initial 
conditions'problem. Indeed, since quintessence and matter energy densities 
evolve at different rates with the universe's expansion, conditions in the 
early universe must be set very carefully in order for them to be comparable 
to the ones existing today. In a tracker model, a very wide range of initial 
conditions yields to a common evolution for today and so the problem can be 
avoided.}. This is why solution admitting tracker behavior leads to the most 
favoured models. As they are characterized by an approximatively constant {\em 
w}, in what follows, we shall focus on equations of state with a constant {\em 
w}.

There are several reasons that lead us to favour a scalar field candidate. 
First of all, while the cosmological constant does not yet possess a 
completely satisfactory physical interpretation, the scalar field appears 
naturally in the field equations of a large number of alternative theories to
general relativity. Moreover, in some of these alternative theories (e.g. 
superstring theory), the scalar terms play an important physical role and  
consequently cannot be neglected. 

In this letter, we shall consider the hypothesis of a constant equation of state 
for quintessence and show that, in this case, the set of field equations and 
conservation laws does not allow to use any potential form. 

\section{The constraint on the potential form}

The field equations of a FLRW spacetime filled with ordinary matter non-coupled 
with a homogeneous scalar field are
\begin{eqnarray} 
\frac{\stackrel{\cdot}{R}^2}{R^2} + \frac{k}{R^2} & = & \frac{1}{3}\,\kappa^2 
\rho_\phi + \frac{1}{3}\,\kappa^2 \rho_M
\label{field1} \\
2\,\frac{\stackrel{\cdot\cdot}{R}}{R} + \frac{\stackrel{\cdot}{R}^2}{R^2}
+ \frac{k}{R^2} & = & - \kappa^2 \,p_\phi 
\label{field2} \\
\stackrel{\cdot\cdot}{\phi} + 3\,\stackrel{\cdot}{\phi}\,
\frac{\stackrel{\cdot}{R}}{R} & = & - V'
\label{field3}
\end{eqnarray}
where we have defined
\begin{eqnarray} \displaystyle
\kappa^2\, \rho_\phi & \equiv & \frac{1}{2}\,\stackrel{\cdot}{\phi}^2 + \,V(\phi) 
\label{density} \\
\kappa^2\,p_\phi & \equiv & \frac{1}{2}\,\stackrel{\cdot}{\phi}^2 - \,V(\phi)
\label{pressure}
\end{eqnarray}
and where we have taken as equation of state for the ordinary fluid: 
$p_M = 0$. In all our equations, the dot denotes the derivative with respect 
to the time coordinate and the prime, the derivative with respect to the 
scalar field. 

The fundamental assumption at the basis of the quintessential hypothesis is 
to consider that the scalar field behaves like a fluid with as 
equation of state
\begin{equation} \displaystyle
p_\phi = w\,\rho_\phi
\label{quintessence}
\end{equation}
where {\em w} is lying between $-1$ and $0$, the limit $w=-1$ corresponding
to the cosmological constant. 

As there is no interaction between the matter field and the scalar field, we 
have to impose the conservation law on these two fields separately: 
\begin{equation} \displaystyle
\kappa^2 \rho_M = \kappa^2\,\rho_{M,0}\,\left(\frac{R_0}{R}\right)^{3} 
\hspace*{10mm}
\kappa^2 \rho_\phi = \kappa^2\,\rho_{\phi,0}\,\left(
\frac{R_0}{R}\right)^{3(1+w)} 
\label{conservphi}
\end{equation}
where the subscript ``$0$'' means ``the current value''. The supernovae
observations being given in terms of the density parameters, it is convenient 
to introduce the following dimensionless quantities:
\begin{equation} \displaystyle
\Omega_k \equiv \frac{- k}{R^2_0\,H^2_0} \hspace*{13mm}
\Omega_\phi \equiv \frac{\kappa^2\,\rho_{\phi,0}}{3\,H^2_0} \hspace*{13mm} 
\Omega_M \equiv \frac{\kappa^2\,\rho_{M,0}}{3\,H^2_0} 
\label{omega}
\end{equation}
constrained by $\Omega_0 \equiv 1 - \Omega_k = \Omega_M + \Omega_\phi$. 
Introducing the definitions (\ref {quintessence})-(\ref {omega}) in the 
field equation (\ref {field1}), we obtain the following differential equation 
for the scale factor $R(t)$:
\begin{equation} \displaystyle
\stackrel{\cdot}{R}^2\, = R_0^2\,H^2_0\,\left[
\Omega_\phi\,\left(\frac{R_0}{R}\right)^{3w+1} + \Omega_M\,\frac{R_0}{R}
+ \Omega_k \right]
\label{firstintegral}
\end{equation}

We shall now transform this relation in a differential equation for the
scalar potential $V(\phi)$. Using combinations of
eqs.(\ref{density})-(\ref{conservphi}), we find the relation between $R(t)$
and $V(\phi)$:
\begin{equation}
\frac{R}{R_0} = \left( \frac{V}{V_0} \right)^{\frac{-1}{3(w+1)}}
\label{rien1}
\end{equation}
where $V_0$, defined by $V_0 \equiv 3\,(1-w)\,H^2_0\,\Omega_\phi\,/\,2$,
represents the current value for the scalar potential $V(\phi)$. The
derivative of (\ref {rien1}) with respect to $\phi$ leads to an expression
of $\stackrel{\cdot}{R}(t)$ as a function of $V(\phi)$ and $V'(\phi)$:
\begin{equation}
\frac{\stackrel{\cdot}{R}}{R_0\,H_0} =
\left( \frac{V}{V_0} \right)^{- \frac{3w+5}{6(w+1)}} \frac{V'}{V'_0}
\label{rien2}
\end{equation}
where $V'_0$, defined by $V'_0 \equiv \pm\,3\,H_0\,\sqrt{(1-w^2)\,V_0\,/\,2}$,
is the current value of $V'(\phi)$. Using (\ref {rien1}) and (\ref {rien2}),
we can write relation (\ref {firstintegral}) in terms of $V(\phi)$ and
$V'(\phi)$ and find the following constraint on the scalar potential:
\begin{equation} \displaystyle
\frac{V'}{V'_0} = \,\sqrt{\Omega_\phi\,\left(\frac{V}{V_0}\right)^2 + 
\Omega_M\,\left( \frac{V}{V_0}\right)^{\frac{w+2}{w+1}} + 
\Omega_k\,\left( \frac{V}{V_0}\right)^{\frac{3w+5}{3(w+1)}} 
}
\label{constraint}
\end{equation}
This relation has been
found assuming that $w \not= -1$. For $w=-1$, we are in the case of the 
cosmological constant which implies that the scalar field and its potential 
are constant, so that eq.(\ref {constraint}) loses its meaning. As we can see 
from (\ref {constraint}), any form of potential $V(\phi)$ with any value of 
{\em w} is not consistent with the field equations and the conservation laws. 

We have been able to solve the constraint (\ref {constraint}) only in some
peculiar cases: 
\begin{enumerate}
\item[1.] \underline{For $k = 0$ ($\forall w$)} (flat FLRW model): 
\begin{equation}
V(\phi) = V_0\,\left[\,\sqrt{\frac{\Omega_M}{\Omega_\phi}}\,sinh\left[
\pm\,\beta_0\,(\phi-\phi_0)+\alpha_0 \right]\, \right]^{2\,(w+1)/w}
\end{equation}
with $\alpha_0 \equiv arcsinh\left[ \sqrt{\Omega_\phi \,/\, \Omega_M}
\right]$ and $\displaystyle \beta_0 \equiv \frac{w}{2}\,\sqrt{3\,/\,(w+1)}$. 
Quintessence model with this potential has already been considered by Chimento
and Jakubi in \cite{chimento} and more recently by Ureno-Lopez and Matos in 
\cite{matos}.
\item[2.] \underline{For $w = -2/3$:} 
{\small
\begin{equation} \displaystyle
V(\phi) = V_0\,\left[\sqrt{\frac{\Omega_M}{\Omega_\phi}}\,
sinh\left[\pm\,(\phi-\phi_0) + \delta_0\right] + \frac{1}{4}\,
\frac{\Omega_k}{\Omega_\phi}
\left(\frac{\Omega_k}{\Omega_M} e^{\mp\,(\phi-\phi_0)-\sigma_0}
- 2\right) \right]^{-1}
\end{equation}
}
whith $e^{\sigma_0} = \left(2\,\Omega_\phi + 2\,\sqrt{\Omega_\phi} 
+ \Omega_k\right)\,/\,\Omega_M$ and $\delta_0 = \sigma_0 + 
ln\sqrt{(\Omega_M)\,/\,(4\,\Omega_\phi)}$. 
\item[3.] \underline{For $w = -1/3$:}\\
\begin{itemize}
\item For $\Omega_\phi + \Omega_k = 0$:
\begin{equation} \displaystyle
V(\phi) = V_0 \left( \frac{\phi_0}{\phi}\right)^4
\end{equation}  
\item For $\Omega_\phi + \Omega_k \not= 0$, it is
\begin{equation} \displaystyle
V(\phi) = V_0\,\left[\sqrt{\frac{\Omega_M}{\Omega_\phi + \Omega_k}}\,
sinh\left[ \pm\,\epsilon_0\,(\phi-\phi_0) + \nu_0 \right]\right]^{-4}
\end{equation}
\noindent where $\epsilon_0 \equiv \sqrt{\left(\Omega_\phi + \Omega_k\right)
\,/\,\left(8\,\Omega_M\right)}$\,\, and $\nu_0 \equiv 
arcsinh\left[2\,\sqrt{2}\,\epsilon_0\right]$. Note that this potential that may 
mimic a negative spatial curvature of the Universe was introduced e.g. by 
Starobinsky in \cite{rien}.
\end{itemize}
\end{enumerate}
In these three cases, $\phi_0$ is the current value of the scalar field $\phi(t)$ 
so that we can always write $V(\phi_0) = V_0$. 

\section{The potentials used in the literature} 

We shall now consider the three most used potential forms in the 
quin\-tes\-sen\-tial literature, namely the inverse power-law form 
$V(\phi) = V_0\,\left(\frac{\phi_0}{\phi}\right)^{a_0}$ (with $a_0 > 1$) 
suggested by SUSY models \cite{potential1}, the exponential form 
$V(\phi) = V_0\,e^{\phi-\phi_0}$ invoqued in the context of Kaluza-Klein or
superstring theories \cite{inflation} and the cosine
form $V(\phi) = \frac{1}{2}\,V_0\,
\left[ cos(\left[\phi - \phi_0 \right]/f) + 1 \right]$
motivated by the physics of pseudo-Nambu-Goldstone bosons \cite{potential2}.

The introduction of the inverse power-law potential in the constraint
(\ref {constraint}) gives 
\begin{equation} \displaystyle
\frac{a_0^2\,V_0^2}{\phi_0^2\,{V'_0}^2}\,\left(\frac{\phi_0}{\phi}
\right)^{2\,(a_0 + 1)} = 
\Omega_\phi \left(\frac{\phi_0}{\phi}\right)^{2\,a_0} + 
\Omega_M \,\left( \frac{\phi_0}{\phi} \right)^{\frac{w+2}{w+1}\,a_0} + 
\Omega_k\, \left( \frac{\phi_0}{\phi} \right)^{\frac{3w+5}{3(w+1)}\,a_0}
\end{equation}
The only way to satisfy this relation is to identify the terms that can have 
the same exponent of $(\phi_0 / \phi)$. In this case, we have two 
possibilities for those identifications:
\begin{eqnarray} \displaystyle 
2\,(a_0 + 1) & = & a_0\,(3\,w + 5)/ 3 (w + 1) \\ 
a_0^2\,V_0^2\,/\,\phi_0^2\,{V'}_0^2 & = & \Omega_k\\
2\,a_0 & = & a_0\,(w + 2)/(w + 1) \label{rien} \\
\Omega_\phi + \Omega_M & = & 0
\end{eqnarray}
or
\begin{eqnarray} \displaystyle 
2\,(a_0 + 1) & = & a_0\,(w + 2)/(w + 1) \label{ident1}\\ 
a_0^2\,V_0^2\,/\,\phi_0^2\,{V'}_0^2 & = & \Omega_M\\  
2\,a_0 & = & a_0\,(3\,w + 5)/ 3 (w + 1) \label{ident2} \\
\Omega_\phi + \Omega_M & = & 0  
\end{eqnarray}

It is easy to see from (\ref{rien}) that no value of $w$ allows one to 
satisfy the first set of relations. The second set can be satisfied only 
for one value of {\em w}: $w = -1/3$. We also find the following constraints
on the other constants:
$a_0 = 4$, $\Omega_M = 1$ and $\Omega_\phi = -\Omega_k = \phi_0^2\,/\,8$. The
negative value of $\Omega_k$ means that the corresponding universe model is
closed ($k=1$). Note also that the value $a_0 = 4$ has already often been 
considered in the context of quintessence \cite{potential1}. However, as it
is well known, the value $\Omega_M = 1$ found in this model is completely
incompatible with those deduced from primordial nucleosynthesis and
supernovae observations ($\Omega_M \approx 0.25$) which leads us to admit
that an inverse power-law potential as we have used is not coherent with 
constraint (\ref {constraint}). 

With the exponential potential, constraint (\ref {constraint})
becomes
\begin{equation}
\left( \frac{V_0}{V'_0} \right)^2 \left( \frac{V}{V_0} \right)^2 =
\Omega_\phi \left( \frac{V}{V_0} \right)^2 +
\Omega_M \left( \frac{V}{V_0} \right)^{\frac{w+2}{w+1}} +
\Omega_k \left( \frac{V}{V_0} \right)^{\frac{3w+5}{3(w+1)}}
\end{equation}
while with the cosine form, it can be written as
\begin{equation}
\left( \frac{V_0}{V'_0} \right)^2 \left[ \frac{V}{V_0} - \left(
\frac{V}{V_0} \right)^2 \right] =
\Omega_\phi \left( \frac{V}{V_0} \right)^2 +
\Omega_M \left( \frac{V}{V_0} \right)^{\frac{w+2}{w+1}} +
\Omega_k \left( \frac{V}{V_0} \right)^{\frac{3w+5}{3(w+1)}}
\end{equation}
By trying to make similar identifications, one can easily see that there
is no value of $w$ able to satisfy the constraint in those two cases. So we
are led again to the same conclusion as in the
first subcase of the inverse power-law potential: all these scalar potential
forms are not consistent with the set of field and conservation equations
issued from quintessential hypothesis.

\section{Conclusion}
In this letter, we have attempted to establish that the field equations and 
the conservation laws strongly constrain the scalar potential form invoked
in the framework of the quintessence. As an application, we have
studied the three most used quintessential potentials. In all cases, it
appeared that they were incompatible with the constraint we found on the 
potential. This can castsome doubt on the way the quintessential hypothesis 
has been presently
implemented in the theoretical framework. However, as it has already been
mentionned in the introduction, our results have been obtained making the
hypothesis of a constant scalar field equation of state wich is a good
approximation only for a tracker solution. So the case of a variable {\em w}
deserves further theoretical investigations \cite{elisa}. As suggested
by Huterer and Turner \cite{concl}, it should be soon possible to
discriminate between a constant and a time-varying {\em w} using future SNeIa
observations and also high precision measurements of the multipole power
spectrum expected from the MAP and Planck Surveyor satellites.
\vspace*{4mm} \\ \noindent
{\large \bf{Acknowledgments}}
\vspace*{2mm} \\ \noindent
This work was supported in part by Belgian Interuniversity Attraction Pole
P4/05 as well as by a grant from {\em Fonds National de la Recherche
Scientifique}. The first manuscript of this letter was written before the very
untimely passing away of Jacques Demaret, to whose this publication is of 
course dedicated.
\end{document}